\documentstyle{mn}
\input{epsf}
\begin{document}
 

\def\g{$\gamma$}
\def\zm{z_{\rm max}}
\def\nh{N_{\rm H}}
\def\nlo{\langle n L\rangle_0}
\def\af{A_{\rm Fe}}
\def\ef{E_{\rm Fe}}
\def\ife{I_{\rm Fe}}
\def\sf{\sigma_{\rm Fe}}
\def\taut{\tau_{\rm T}}
\def\ec{E_{\rm c}}
\def\ginga{{\it Ginga}}
\def\asca{{\it ASCA}}
\def\exosat{{\it EXOSAT}}
\def\iue{{\it IUE}}
\def\hst{{\it HST}}
\def\luv{L_{\rm UV}}
\def\fuv{F_{\rm UV}}
\def\le{L_{\rm E}}
\def\lxg{L_{{\rm X}\gamma}}
\def\granat{{\it GRANAT}}
\def\as{\alpha_{\rm s}}
\def\is{I_{\rm s}}
\def\rosat{{\it ROSAT}}
\def\ee{e$^\pm$}
\def\rs{r_{\rm S}} 
\def\rxg{r_{{\rm X}\gamma}}
\def\lnt{\ell_{\rm nth}}

\hyphenation{Max-well-ian brems-strahl-ung syn-chro-tron
black-body ap-pen-dix i-so-tro-pic com-pact-ness com-pact-nes-ses}
 
\def\ast{\mathchar"2203} \mathcode`*="002A
\newbox\grsign \setbox\grsign=\hbox{$>$} \newdimen\grdimen \grdimen=\ht\grsign
\newbox\simlessbox \newbox\simgreatbox \newbox\simpropbox
\setbox\simgreatbox=\hbox{\raise.5ex\hbox{$>$}\llap
     {\lower.5ex\hbox{$\sim$}}}\ht1=\grdimen\dp1=0pt
\setbox\simlessbox=\hbox{\raise.5ex\hbox{$<$}\llap
     {\lower.5ex\hbox{$\sim$}}}\ht2=\grdimen\dp2=0pt
\setbox\simpropbox=\hbox{\raise.5ex\hbox{$\propto$}\llap
     {\lower.5ex\hbox{$\sim$}}}\ht2=\grdimen\dp2=0pt
\def\simgreat{\mathrel{\copy\simgreatbox}}
\def\simless{\mathrel{\copy\simlessbox}}
\def\simprop{\mathrel{\copy\simpropbox}}

\topmargin=-1cm
 
\title[Broad-band \g-ray and X-ray spectra of NGC 4151]
{Broad-band gamma-ray and X-ray spectra of NGC 4151 
and their implications for physical processes and geometry}

\author[A. A. Zdziarski, W. N. Johnson and P. Magdziarz]
{\parbox[]{6.in} {Andrzej A. Zdziarski$^1$, W. Neil Johnson$^2$ and 
Pawe\l\ Magdziarz$^3$} \\
$^1$Copernicus Astronomical Center, Bartycka 18, 00-716 Warsaw, Poland,
Internet: aaz@camk.edu.pl\\
$^2$E. O. Hulburt Center for Space Research,
Naval Research Lab, Washington, DC 20375, USA\\
$^3$Astronomical Observatory, Jagiellonian University, Orla 171, 30-244 
Cracow, Poland, Internet: pavel@camk.edu.pl\\} 
 

\maketitle
\begin{abstract}

We study \g-ray observations of NGC 4151 by {\it GRO}/OSSE contemporaneous 
with X-ray observations by \rosat\/ and \ginga\/ in 1991 June and with \asca\/ 
in 1993 May. The spectra are well modeled by thermal Comptonization and a dual 
neutral absorber. We also find, for the first time for NGC 4151, a 
Compton-reflection spectral component in the \ginga/OSSE data. When reflection 
is taken into account, the intrinsic X-ray energy spectral index is $\alpha 
\sim 0.8$ and the plasma temperature is $\sim 60$ keV for both observations, 
conditions which imply an optical depth of $\sim 1$. The X-ray spectral index 
is within the range, $\alpha\simeq 0.95\pm 0.15$, observed from other Seyfert 
1s. Also, the OSSE spectra of those and other observations of NGC 4151 are 
statistically undistinguishable from the average OSSE spectrum of radio-quiet 
Seyfert 1s. Thus, NGC 4151 observed in 1991 and 1993 has the intrinsic 
X-ray/\g-ray spectrum typical for Seyfert 1s, and the main property 
distinguishing it from other Seyfert 1s is a large absorbing column of $\sim 
10^{23}$ cm$^{-2}$. We find no evidence for a strong, broad and redshifted, Fe 
K$\alpha$ line component in the \asca\/ spectrum of 1993 May. Also, the 
Compton-reflection component in the \ginga/OSSE spectrum is a few times too 
small to account for the strength of the broad/redshifted line reported 
elsewhere to be found in this and other \asca\/ spectra of NGC 4151. 

On the other hand, we confirm previous studies in that archival X-ray data do 
imply strong intrinsic X-ray variability and hardness of the intrinsic 
spectrum in low X-ray states. An observed softening of the intrinsic X-ray 
spectrum with the increasing flux implies variability in \g-rays weaker than 
in X-rays, which agrees with the 100 keV flux changing only within a factor of 
2 in archival OSSE and \granat/SIGMA observations. 

The relative hardness of the intrinsic X-ray spectrum rules out the 
homogeneous hot corona/cold disk model for this source. Instead, the hot 
plasma has to subtend a small solid angle as seen from the source of UV 
radiation. If the hot plasma is purely thermal, it consists of electrons 
rather than \ee\ pairs. On the other hand, the plasma can be pair-dominated if
a small fraction of the power is nonthermal. 

\end{abstract} 

 
\section{INTRODUCTION}
\label{s:intro}

NGC 4151 is a nearby Seyfert 1.5 galaxy at $z=0.0033$. In spite of its 
proximity and the wealth of X-ray and \g-ray data accumulated over the last 20 
years, the nature of its nucleus remains poorly understood. Although it is the 
brightest Seyfert in hard X-rays it appeared to be distinctly different from 
average Seyfert 1s. Its X-ray spectrum is highly variable in the 2--10 keV 
energy spectral index, $\alpha\sim 0.3$--0.8 (e.g.\ Yaqoob \& Warwick 1991, 
hereafter YW91; Yaqoob et al.\ 1993, herafter Y93). This hardness of the X-ray 
spectrum contrasts typical Seyfert 1s, which have $\alpha \simeq 0.95\pm 0.15$ 
on average (Nandra \& Pounds 1994). Furthermore, no characteristic spectral 
upturn above $\sim 10$ keV due to Compton reflection from cold matter, typical 
for Seyfert 1s (Nandra \& Pounds 1994), was found in X-ray spectra of NGC 4151 
from {\it Ginga\/} (Y93). 

In this work, we present and discuss results of the monitoring of NGC 4151 by 
the OSSE detector aboard {\it GRO\/} from 1991 to 1993. In 1991 June and 1993 
May, the OSSE data can be supplemented by data in X-rays from {\it ROSAT}/{\it 
Ginga}, and {\it ASCA}, respectively. These data suggest the intrinisic 
X-ray/\g-ray (hereafter abbreviated as X\g) spectrum which is relatively 
steady in both the shape and amplitude, in contrast with many earlier 
observations in X-rays. We also study the presence of a Compton-reflection 
component and the form of the Fe K$\alpha$ line in the data. We also consider 
implications of archival X\g\ data from \exosat, \ginga, and {\it GRANAT}. 

After presenting the data, we consider their implications for physical 
processes in NGC 4151. We study Comptonization in the X\g\ source as well the 
presence of \ee\ pairs and nonthermal acceleration. From that, we obtain 
strong constraints on the parameters and geometry of the central region of NGC 
4151.

\section{X-RAY/GAMMA-RAY SPECTRA}
\label{s:obs}

We consider first two broad-band X\g\ spectra combining observations by OSSE 
in 1991 June/July and 1993 May with contemporaneous observations in X-rays by 
\ginga/\rosat\/ and \asca, respectively. Then we consider other available data 
from OSSE, \ginga, \exosat, and \granat. 

\subsection{Fitted Models}
\label{ss:model}

We use {\sc xspec} v.\ 9.0 (Arnaud 1996) for spectral fitting. The quoted 
errors are for 90 per cent confidence limit based on a $\Delta \chi^2=2.7$ 
criterion (Lampton, Margon \& Bowyer 1976). Note that this confidence range 
correspond to the probability that a fitted parameter is in this range 
regardless of the values of all other parameters of a model (e.g.\ Press et 
al.\ 1992). Model parameters are given at $z=0.0033$. 

We use thermal Comptonization of Lightman \& Zdziarski (1987, hereafter LZ87) 
as the hard continuum model (see Appendix). The parameters of the model are 
the plasma temperature, $kT$, and the spectral index, $\alpha$, of the 
low-energy asymptotic power law. The second parameter is used instead of the 
geometry-dependent Thomson optical depth, $\tau$, of the plasma. As an 
alternative form of the continuum, we use a power-law spectrum with an 
exponential cutoff. 

We allow for the presence of a Compton-reflection component in the spectrum, 
which arises when cold matter (e.g.\ an accretion disk) subtends a substantial 
solid angle as seen from the X\g\ source (Lightman \& White 1988). We use 
angle-dependent reflection Green's functions of Magdziarz \& Zdziarski (1995) 
and allow the reflecting medium to be either neutral or ionized and with the 
same abundances as in the absorber (see below). The relative contribution of 
reflection is measured by the ratio, $R$, of the flux emitted towards the 
reflector to that emitted outward. Equivalently, $2\pi R$ gives the solid 
angle covered by the reflector as seen from the X\g\ source, with $R=1$ 
corresponding to an isotropic continuum source above a slab. 

The continuum is modified by absorption. We use here a dual absorber model, 
i.e., the product of complete absorption (with the hydrogen column density 
$N_2$) and partial covering by neutral medium (with the hydrogen column $N_1$
and the covering factor $C_{\rm f}$; e.g.\ Weaver et al.\ 1994, hereafter 
W94). We use the opacities of Ba\l uci\'nska-Church \& McCammon (1992) and the 
abundances of Anders \& Ebihara (1982). However, the ratio of the Fe abundance 
to that of Anders \& Ebihara (1982), $\af$, is a free parameter.

As an alternative, we also consider an ionized (`warm') absorber (Halpern 
1984). We use a model of Done et al.\ (1992) but with the abundances of Anders 
\& Ebihara (1982) instead of those of Lang (1974). Also, the iron edge 
energies are corrected (P. \.Zycki, private communication) from the 
approximate values of Reilman \& Manson (1979) to those of Kaastra \& Mewe 
(1993). The former authors employ the Hartree-Slater approximation, which is 
accurate to a few per cent only (e.g.\ their K-edge energy of neutral Fe is 
6.9 keV instead of 7.1 keV). The corrected absorber opacities agree then with 
those of Ba\l uci\'nska-Church \& McCammon (1992) in the limit of zero 
ionization. We assume the absorber temperature of $10^5$ K (Krolik \& Kallman 
1984). The ionization parameter is defined by $\xi=L/(nr^2)$, where $L$ is the 
5 eV--20 keV luminosity in an incident power-law spectrum and $n$ is the 
density of the absorber located at distance $r$ from the illuminating source. 

NGC 4151 exhibits a strong soft excess component, whose complex nature is not 
fully understood (e.g.\ Warwick, Done \& Smith, hereafter WDS95). We model it 
simply as a power law cut off at an $e$-folding energy, $E_{\rm s}$. Its value 
is fixed at $E_{\rm s}=2$ keV, which provides good fits for both 1991 June and 
1993 May data sets. The index of the soft power law, $\as$, is allowed to be 
different from that in the hard X-rays, which was found necessary by WDS95. 
This difference rules out the origin of the soft excess solely due to 
scattering by warm electrons external to the X\g\ source, which origin was 
proposed by W94. The soft component is absorbed by a neutral medium with the 
Galactic column density, $N_0=2.1\times 10^{20}$ cm$^{-2}$ (Stark et al.\ 
1992), except for the 1991 June observation. 

NGC 4151 also emits an Fe K$\alpha$ line around 6.4 keV. We model it here 
as a Gaussian ($f_{\rm Fe}$) centered at $\ef$ with a width of $\sf$, the 
total photon flux $I_{\rm Fe}$, and absorbed in the same way as the continuum.  

For the dual absorber, the model has the form, 
 \begin{eqnarray}
\lefteqn{
F(E)=e^{-\sigma N_0} \left\{e^{-E/E_{\rm s}} I_{\rm s} 
E^{-1-\as} +\right.} \nonumber \\
&& \left. \left(1-C_{\rm f} +C_{\rm f} e^{-\sigma N_1}\right) e^{-\sigma N_2}
\left[I_{\rm Fe} f_{\rm Fe}(E) +I_{\rm c} f_{\rm c}(E)\right] 
\right\},
\label{eq:model}
\end{eqnarray}
 where $E$ is the photon energy in keV, $f_{\rm c}$ represents the hard 
continuum including reflection, $I_{\rm c}$ is its 1 keV normalization, and 
$\sigma$ is the bound-free cross section of neutral matter. In the case of 
ionized absorber, the two first factors in the second line of equation 
(\ref{eq:model}) are replaced by $e^{-\sigma_{\rm ion} N_1}$, where 
$\sigma_{\rm ion}$ is the bound-free cross section of ionized matter with the 
hydrogen column $N_1$. 

\subsection{June 1991} 
\label{ss:91}

\subsubsection{The data}
\label{sss:91data}

NGC 4151 was observed by OSSE 1991 June 29--July 12. The observation was
reported before by Maisack et al.\ (1993). The data used here have been 
obtained with the response matrix and calibration revised since then (see 
Johnson et al.\ 1996). The response revision has resulted in the $\sim 60$--70 
keV flux being about 10 per cent larger now. Also, the 50--60 keV channel data 
have been added. 

The present data include estimated systematic errors.  These systematics were 
computed from the uncertainties in the low energy calibration and response of 
the detectors using both in-orbit and prelaunch calibration data.  The 
energy-dependent systematic error was estimated at a value of 3 times the 
computed uncertainty and was added in quadrature to the statistical errors 
prior to spectral fitting. For NGC 4151, this systematic error is $\sim 12$ 
per cent of the statistical error at 50 keV and decreases to less than 0.1 per 
cent above 130 keV. 

Close in time to the OSSE observation, NGC 4151 was observed by {\it Ginga\/}, 
1991 May 31--June 2 (Y93), and by {\it ROSAT}, 1991 May 31--June 1. The 
\ginga\/ (top-layer only) and \rosat\/ data have been analyzed by WDS95. Here 
we use both the mid-layer and top-layer \ginga\/ data (Turner et al.\ 1989), 
as recently calibrated for the Leicester \ginga\/ database (D. Smith, private 
communication). The mid-layer and top-layer data are fully consistent with 
each other above 9 keV, and the mid-layer data are more accurate than the 
top-layer ones at $\simgreat 15$ keV. We use \ginga\/ data from a time 
interval overalapping with the \rosat\/ observation (see WDS95). Note that 
this subset differs from the {\it Ginga\/} spectrum `b' (in the notation of 
Y93), which was used in previous fits by Zdziarski, Lightman \& 
Macio{\l}ek-Nied\'zwiecki (1993, hereafter ZLM93) and Titarchuk \& 
Mastichiadis (1994, hereafter TM94). In order to allow for residual 
calibration uncertainties, we have added a 0.5 per cent systematic error to 
the \ginga\/ data (Turner et al.\ 1989), and a 2 per cent error to the 
\rosat\/ data (WDS95). We then fit jointly the 0.2--2 keV \rosat\/ data, the 
2--22 keV \ginga\/ top-layer data, the 9--24 keV \ginga\/ mid-layer data, and 
the 50--1000 keV OSSE data. 

In the models below, we take into account the contribution to the {\it 
Ginga\/} spectrum from the BL Lac 1E1207.9+3945, which is located $\sim 5$ arc 
min from the nucleus of NGC 4151 and is not resolved by {\it Ginga\/}. The BL 
Lac is resolved by {\it ROSAT\/} (WDS95), which observation yields a power law 
spectrum with $\alpha\simeq 1.1$ and a normalization implying an about 10 per 
cent contribution to the {\it Ginga\/} spectrum at 2 keV. (We arbitrarily 
assume the BL Lac power law is cut off exponentially with an $e$-folding 
energy of 300 keV; the value of this energy has negligible effect on our 
fits.) 

\subsubsection{Results}
\label{sss:91res}

Our baseline model contains a hard continuum due to thermal Comptonization, 
a soft X-ray component, a dual absorber, and a K$\alpha$ line with the energy 
and width of 6.4 keV and 0.1 keV, respectively (see Section \ref{ss:model}). 
We first fix the Fe abundance at the solar value of Anders \& Ebihara (1982). 
We obtain an inacceptable fit with $\chi^2=194/126$ d.o.f. There are very 
strong residuals in the 5--8 keV range, which form an apparent broad line 
between 5 and 6 keV and an edge around 7 keV, as shown in Fig.\ 
\ref{fig:res91}{\it a}. However, these systematic residuals disappear when the 
Fe abundance is allowed to vary (reaching $\af= 2.8^{+0.3}_{-0.3}$; cf.\ YW91; 
Y93), as shown in Fig.\ \ref{fig:res91}{\it b}. The reduction in $\chi^2$ is 
very large, to $\chi^2= 99/125$ (at the continuum parameters of 
$\alpha=0.74_{-0.01}^{+0.02}$, $kT=54^{+9}_{-6}$ keV). The residuals at $\af= 
1$ are explained by the presence of a strong K edge in the data, which causes 
the fitted local continuum to soften in order to account for the dip above the 
edge energy. This in turn gives rise to positive residuals below the edge and 
negative ones above it. The residual at $\sim 6.5$ keV is close to null 
because of compensation by an enhanced flux in the narrow $\sim 6.4$ keV line 
($\sim 4$ times stronger at $\af=1$ than at free $\af$). Then the residuals 
between $\sim 4$ and 6.5 keV form a broad, line-like, feature.  

\begin{figure}
\begin{center}
\leavevmode
\epsfxsize=8.4cm \epsfbox{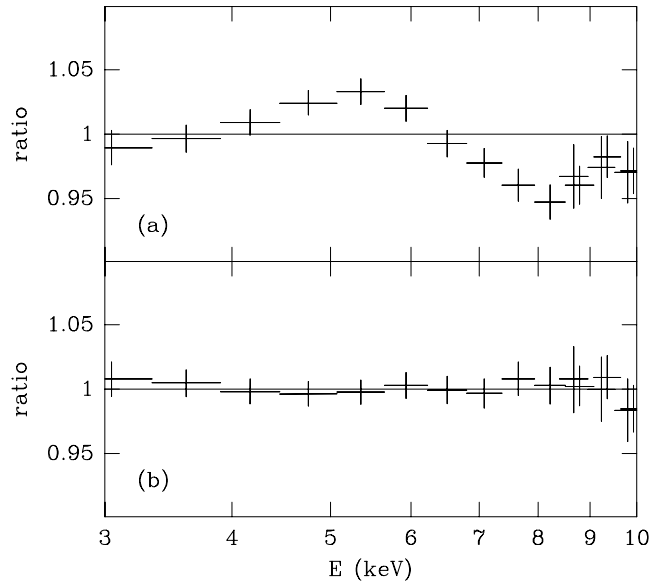}
\end{center}
\caption{The 3--10 keV residuals to fits to 1991 June \ginga\/ data. {\it 
(a)\/} The Fe abundance of Anders \& Ebihara (1982). {\it (b)\/} The best fit 
Fe abundance. The broad $\sim 5.5$ keV feature apparent at the solar Fe 
abundance disappears now. See text. 
 }
\label{fig:res91}
\end{figure} 

In order to test the uniqueness of the obtained continuum shape, we also fit a 
power law with an exponential cutoff. We obtain a much worse fit, with 
$\chi^2=112/125$ d.o.f., for $\alpha=0.56$, the $e$-folding energy of 
$\ec=121$ keV, and $\af=3.5$. The model fits both \ginga\/ and OSSE data worse 
than the thermal-Compton model. (The value of $\alpha$ is lower than that for 
the Comptonization model because of the different shapes of high-energy 
cutoffs of the two models fitted to the OSSE data.) Since the cut-off 
power-law model both gives a worse fit as well as it does not correspond to 
any physical process for the obtained parameters (see Section \ref{ss:par}), 
we do not consider it any more. 

We then compare our fits with the dual neutral absorber to those with a warm 
absorber and a soft X-ray component of the same form as for the dual absorber 
(see Section \ref{ss:model}). We use thermal Comptonization for the hard 
continuum and allow for variable $\af$. We obtain a fit with a harder 
intrinsic power law and much worse statistically than that for the dual 
absorber; $\chi^2=155/126$ d.o.f.\ at $\af=1.5$ with 4--10 keV residuals 
similar to those for the dual absorber at $\af=1$. The warm absorber also 
gives a bad fit to the \asca\/ data set of 1993 May.  These results support 
those of Warwick et al.\ (1996), who found that the time-variability pattern 
of soft and hard X-ray variability are difficult to reconcile with the ionized 
absorber model. Also, Kriss et al.\ (1995) have found from HUT data that the 
UV absorber consists of both high and low-ionization gas in the line of sight, 
which rules out the simple model of WDS95 with the UV and X-ray absorbers 
being the same and the low neutral H column in the UV absorber explained 
solely by strong photo-ionization of the X-ray absorber. Therefore, we decided 
to use the dual absorber model throughout. We note that a good fit of the warm 
absorber model to the \rosat/\ginga\/ spectrum of NGC 4151 was obtained by 
WDS95 {\it before\/} the Fe ion edge energies have been corrected (see Section 
\ref{ss:model}). 

We then examine the presence of Compton reflection (Section \ref{ss:model}). 
We assume first the inclination of $i=65^\circ$, which is the minimum 
inclination allowed by \hst\/ observations of Evans et al.\ (1993). We find 
that adding a reflection component significantly improves the fit, reducing 
$\chi^2$ by 10.5, to 88.6/124 d.o.f. This corresponds to the probability of 
0.02 per cent that the fit improvement was by chance. The reflection fraction 
is $R= 0.43^{+0.23}_{-0.22}$. The spectrum is shown in Fig.\ 
\ref{fig:spec}{\it a\/} and the fit parameters are given in Table 1. We 
caution, however, that we do not directly see a hardening of the spectral 
slope above $\sim 10$ keV in NGC 4151 due to strong absorption, whereas such 
hardenings are seen in many other, less absorbed, Seyfert 1s (Nandra \& Pounds 
1994). Thus, the evidence for reflection in NGC 4151 is indirect, and the 
improvement of the fit may possibly be an artifact of our choice of the models 
for the continuum and absorber. 

\begin{table*}
\centering
\caption{Fit parameters for 1991 June and 1993 May observations for the 
thermal-Compton model with Compton reflection. $I_{\rm c}$, $I_{\rm s}$ are in 
$10^{-3}$ cm$^{-2}$ s$^{-1}$ keV$^{-1}$, $I_{\rm Fe}$ is in $10^{-5}$ cm$^{-
2}$ s$^{-1}$; $N$ are in $10^{22}$ cm$^{-2}$, and $kT$ is in keV. 
 } 
\begin{tabular}{lccccccccccc}
\hline
Obs.
& $kT$ & $\alpha$ & $R$ & $I_{\rm c}$
& $\alpha_{\rm s}$ & $I_{\rm s}$ 
& $N_1$ & $C_{\rm f}$ & $N_2$ & $\af$
& $I_{\rm Fe}$\\ 
91 
& $88^{+56}_{-25}$ & $0.80^{+0.03}_{-0.03}$ &$0.43^{+0.23}_{-0.22}$ 
&$95^{+6}_{-6}$ 
& $1.4^{+0.1}_{-0.1}$ & $2.5^{+0.1}_{-0.1}$ 
& $7.3^{+1.2}_{-0.8}$ & $0.74^{+0.09}_{-0.11}$ & $4.3^{+0.7}_{-0.7}$ 
& $2.2^{+0.4}_{-0.4}$
& $10.0^{+9.0}_{-9.7}$\\
93 
& $96^{+64}_{-27}$ & $0.77^{+0.03}_{-0.02}$ & 0.43f &$86^{+8}_{-7}$ 
& $0.7^{+0.1}_{-0.1}$ & $3.5^{+0.2}_{-0.1}$ 
& $13.9^{+1.3}_{-1.1}$ & $0.79^{+0.02}_{-0.03}$ 
& $3.8^{+0.4}_{-0.4}$ & $1.3^{+0.3}_{-0.3}$
& $64^{+14}_{-11}$\\
\hline
\end{tabular}
\label{t:all}
\end{table*}

\begin{figure*}
\begin{center}
\leavevmode
\epsfxsize=12.35cm \epsfbox{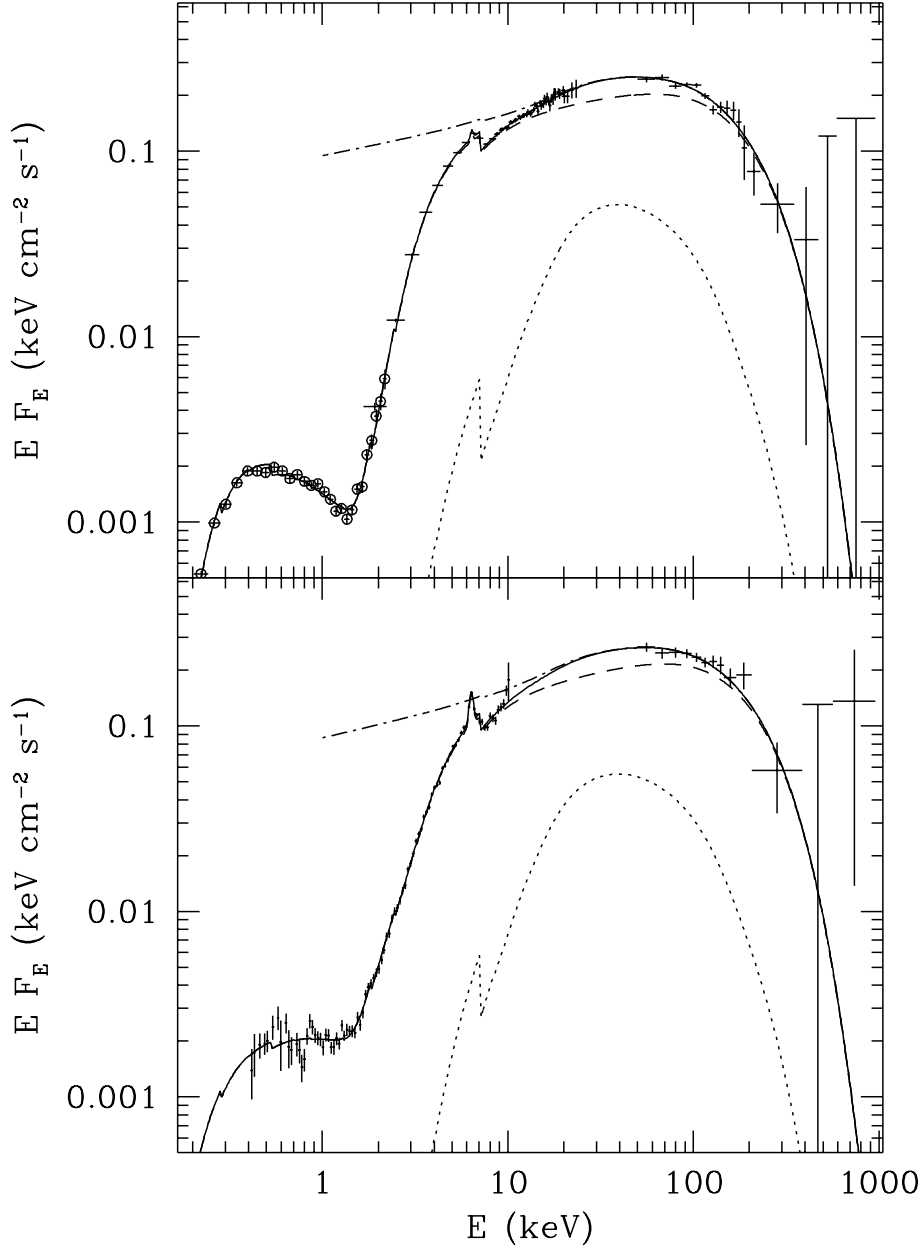}
\end{center}
\caption{The X\g\ spectra of NGC 4151 as observed {\it (a)\/} in 1991 
June/July by \rosat, \ginga\ and OSSE, and {\it (b)\/} in 1993 May by \asca\/ 
and OSSE.  The dashed curves give the absorbed thermal Comptonization spectra, 
the dotted curves, the reflection components, and the solid curves give the 
models of the observed spectra. The dot-dashed curves give the unabsorbed 
continua (without the soft X-ray excess and the K$\alpha$ line). The model 
parameters are given in Table 1. The plotted data have been rebinned for 
clarity, and the upper limits are 2-$\sigma$. The \rosat\/ data are marked 
with circles. The data from all 4 \asca\/ detectors have been co-added for 
plotting. 
 }
\label{fig:spec}
\end{figure*} 

Reflection at $R=0.43$ results in an increase of the fitted $kT$, from 
54 keV to $88^{+56}_{-25}$ keV (but see Section \ref{ss:par}), and an increase 
of the spectral index to $\alpha=0.80^{+0.03}_{-0.03}$. The soft X-ray 
component is absorbed by $N_0=3.3^{+0.3}_{-0.3}\times 10^{20}$ cm$^{-2}$. The 
best-fit Fe abundance is lower when reflection is included, $\af= 
2.2^{+0.4}_{-0.4}$, because a part of the K edge is now accounted for by the 
reflection component.  

We note that the \ginga\/ top-layer data alone are consistent with the 
presence of reflection ($R=0.47$ gives the best fit), but $R=0$ is within the 
90 per cent confidence interval. Addition of both the \ginga\/ mid-layer data 
and the OSSE data strongly narrows that confindence interval. This is due, in 
particular, to the mid-layer data being more accurate than the top-layer data 
above $\sim 15$ keV, where the relative contribution of the reflection 
component peaks.  

We have re-fitted the alternative models of an exponentially cut-off power 
law and of a warm absorber (see above) including now Compton reflection. We
confirm that they provide much worse fits than our baseline thermal-Compton, 
dual-absorber, model even when reflection is included. E.g.\ $\chi^2=102$ for 
the cut-off power-law model with reflection ($\alpha=0.66$, $\ec=150$ keV, 
$R=0.38$) and the dual absorber, i.e., $\Delta\chi^2=13$ with respect to the 
corresponding thermal-Compton model. 

The relative reflection fraction depends on the viewing angle (Magdziarz \& 
Zdziarski 1995). The value of $i=65^\circ$ (Evans et al.\ 1993) is rather 
uncertain. Therefore, we also consider two other inclinations, in particular 
$i=20^\circ$ ($\cos i=0.95$), which is the maximum inlination consistent with 
the K$\alpha$ disk-line fits to \asca\/ data by Yaqoob et al.\ (1995, 
hereafter Y95). Inclinations of $\cos i=0.2$ and 0.95 give $R= 0.76^{+0.40}_{-
0.39}$ and $R= 0.24^{+0.14}_{-0.12}$, respectively, both with the same 
$\chi^2$ as for $i=65^\circ$. Note that the product $R\cos i$ (which 
determines the projected area emitting reprocessed UV flux, see Zdziarski \& 
Magdziarz 1996), increases with $\cos i$; $R\cos i=0.15$, 0.18, 0.23, for 
$\cos i=0.2$, 0.42, and 0.95, respectively. 

We then examine the effect of ionization of the reflector. We model it in the 
same way as ionization of the warm absorber (see Section \ref{ss:model}). We 
obtain $\xi=0^{+65}$ erg s$^{-1}$ cm$^{-1}$, i.e., the reflector is close to 
neutral. Ionization reduces the fitted $R$, which becomes 0.32 only 
($i=65^\circ$) at the upper limit on $\xi$. 

Furthermore, the amount of reflection depends on the overall metal abundance 
(with respect to H and He). Its best fit value equals 1.0 (with the free Fe 
abundance), with the 90 per cent confidence upper limit at 1.9. At this limit, 
the relative reflection is only about 10 per cent larger than for the solar 
abundances, e.g.\ $R=0.27$ at $i=20^\circ$. Thus, varying the overall metal 
abundance has only small effect on the amount of Compton reflection. We have 
also tested the effect of replacing the abundances of Anders \& Ebihara (1982) 
by those of Anders \& Grevesse (1989). The main difference between the two is 
the Fe abundance about 40 per cent higher in the latter. We obtain virtually 
the same spectral parameters and the values of $\chi^2$ for both abundances. 
The Fe abundance with respect to that of Anders \& Grevesse (1989) for the 
model thermal Comptonization and neutral reflection is $\af=1.5^{+0.3}_{-
0.2}$, which is just the range expected from rescaling the corresponding value 
in Table 1.   

A K$\alpha$ line is seen in the X-ray spectrum. Allowing the line width and 
energy to be free parameters, we obtain an insignificant reduction of $\Delta 
\chi^2=- 0.8$ at $\ef=6.2$ keV and $\sf=0$ keV with respect to that at the 
fixed values of 6.4 keV and 0.1 keV, respectively. Thus, the data are 
consistent with the line being narrow and not redshifted. The equivalent width 
(EW) is $34^{+32}_{-33}$ eV (at $\ef=6.4$ keV and $\sf=0.1$ keV and defined 
with respect to the hard continuum only).  The predicted EW of the line from 
reflection at the fitted $R\simeq 0.4\pm 0.2$ (for $i=65^\circ$) is $\sim 
60\pm 30$ eV (George \& Fabian 1991), which then constrains reflection to to 
$R\simless 0.4$. This constraint can be relaxed if the line is broader or the 
reflecting medium is moderately ionized (Ross \& Fabian 1993), which 
possibilities are allowed by the \ginga\/ data. In order to test the former 
possibility we add a second, broad and redshifted, Gaussian component to the 
model (Y95). For $\ef=5.7$ keV and $\sf=0.7$ keV, which are typical values 
obtained by Y95, we obtain $\Delta \chi^2=-0.3$ only, i.e., the data allow but
do not require a second component of the line. The allowed EW of the broad 
line is $30^{+100}_{-30}$ eV, which allows $R$ to be in the range from the 
continuum fit. 

\subsection{May 93}
\label{ss:93}

\subsubsection{The data}
\label{sss:93data}

NGC 4151 was observed by OSSE 1993 May 24--31. During that period, it was also 
observed by \asca\/ on 1993 May 25 (W94). OSSE observed no statistically 
significant variability of the 50--150 keV flux, with the test of the constant 
flux giving $\chi_\nu^2=0.22$ for 7 d.o.f., and the average flux having 2.5 
per cent dispersion. Thus the week-long OSSE spectrum can be used together 
with the single-day \asca\/ spectrum. The OSSE spectrum has been processed in 
the same way as that of 1991 June (see Section \ref{sss:91data}). 

We have extracted the \asca\/ spectrum from the HEASARC archive using the 
\asca\/ software release of 1996 May ({\sc ascaarf} v.\ 2.61). This corrects 
some inaccuracies of the detectors and telescope responses in the previous 
release ({\sc ascaarf} v.\ 2.53).  In particular, the effective area of the 
GIS detectors is now about 20 per cent lower, and an instrumental broad 
spectral feature around 6 keV is now corrected for. 

We used the standard data-screening criteria to select good data (see W94). 
The extraction of SIS data was done in rectangular regions covering parts of 
the chips 1 and 3 for SIS0 and SIS1, respectively. This maximized the usable 
SIS count rate from the observation, which was done in the 4-CCD mode with the 
source off-center (see W94). Still, a relatively large number of counts could 
not be recovered, which significantly reduced the the normalization of the 
obtained SIS spectra and rendered it unreliable (as noted by W94). Therefore, 
we use the GIS spectra (as corrected in the {\sc ascaarf} v.\ 2.61) for the 
absolute normalization. We note that the current GIS spectra yield the 8--10 
keV fluxes about 1.2 times those obtained with the previous \asca\/ software 
(used, e.g.\ in Y95). 

In fitting, we use the 0.4--10 keV SIS data, 0.8--10 keV GIS data, and the 
50--1000 keV OSSE data. The \asca\/ data are rebinned to have at least 20 
counts per bin, as required for the validity of the $\chi^2$ statistics. Each 
of the four \asca\/ data sets is allowed to have free overall normalization, 
and the normalization of the OSSE data is tied to the average normalization of 
GIS2 and GIS3.  Note that the \asca\/ data do not include systematic errors, 
which results in a relatively high $\chi^2_\nu$. 

\subsubsection{Results}
\label{sss:93res}

We use the same models as in Section \ref{ss:91} to the combined \asca/OSSE 
data. We first fit the model with thermal Comptonization, dual absorber, a 
soft X-ray component, free K$\alpha$ line parameters, and no reflection. We 
obtain $\chi^2=1646/1683$ d.o.f. The hard continuum parameters, $\alpha= 
0.72^{+0.03}_{-0.02}$ and $kT=63^{+13}_{-13}$ keV as well its normalization, 
are very similar to the parameters of the corresponding model for the 1991 
June data set (Section \ref{ss:91}). We thus see that the intrinsic state of 
NGC 4151 in 1993 May is very close to that in 1991 June. 

Since the 1991 June data show a Compton reflection component, we include it in 
the present data set as well. The energy range of the \asca\/ data precludes 
an independent determination of the strength of reflection and we simply fix 
it at the best-fit value for the 1991 June data, $R=0.43$. The model yields 
the same $\chi^2=1646/1683$ d.o.f.\ as for $R=0$. The continuum parameters are 
now $\alpha =0.77^{+0.03}_{-0.02}$, $kT=96^{+64}_{-27}$ keV with the 1 keV 
normalization of $I_{\rm c}= 8.6^{+0.8}_{-0.7}\times 10^{-5}$ cm$^{-2}$ s$^{-
1}$, which are indeed the same within the statistical uncertainties as the 
corresponding parameters of the 1991 June observation, see Table 1. The data 
and model are shown in Fig.\ \ref{fig:spec}{\it b}. 

We have also tested a model with the ionized absorber instead of the dual 
neutral absorber, analogously to the 1991 June data (Section \ref{sss:91res}).
We find that model gives a much worse fit, with $\Delta \chi^2=+128$ (at free 
$E_{\rm s}$ and the absorber temperature of $10^5$ K; $\Delta \chi^2=+67$ at 
$10^6$ K). The residuals for the fit resemble those shown in Fig.\ 
\ref{fig:res91}. Thus, ionized absorption does not provide a good model for 
the X-ray data. 

We find that the soft X-ray component in the spectrum of NGC 4151 is also 
consistent with being constant. The 0.5--1 keV flux is $\simeq 2\times 10^{-
12}$ erg cm$^{-2}$ s$^{-1}$ for both 1991 June and 1993 May observations. 
Note that contributions from extended soft X-ray emission (e.g.\ Morse et al.\ 
1995) are different for each of the observations, which accounts for at least 
some of the small differences in the soft X-ray spectra seen in Fig.\ 
\ref{fig:spec}. We also note that although the soft X-ray excess in the 
\rosat/ginga\/ data (Section \ref{ss:91}) can be fitted by bremsstrahlung 
better than by a power law (WDS95), bremsstrahlung at the same temperature 
provides a bad fit to the soft X-ray excess in the \asca\/ data ($\Delta 
\chi^2=+ 58$). On the other hand, the model of equation (1) provides a good 
description of both data sets. 

The two spectra strongly differ, however, in absorption and the strength of 
the K$\alpha$ line. The column density of the partial coverer is now about a 
factor two larger, $N_1 \simeq 1.4\times 10^{23}$ cm$^{-2}$, and the K$\alpha$ 
line is now much stronger, EW $=250^{+50}_{-50}$ eV, than in 1991 June. This 
is suggestive of the difference in the line strength between the two spectra, 
$\Delta$EW $\simeq 200$ eV, being due to emission by the additional absorption 
column in 1993 May (Makishima 1986). The parameters of the line are $\ef 
=6.36^{+0.03}_{-0.04}$ keV, $\sf=0.16^{+0.07}_{-0.09}$ keV. This corresponds 
to the range of cloud velocities of $\Delta v\simeq 18^{+8}_{-10}\times 10^3$ 
km s$^{-1}$ (Lang 1974), which suggest clouds at $\simgreat 10^2$ 
Schwarzschild radii. 

\begin{figure}
\begin{center}
\leavevmode
\epsfxsize=8.4cm \epsfbox{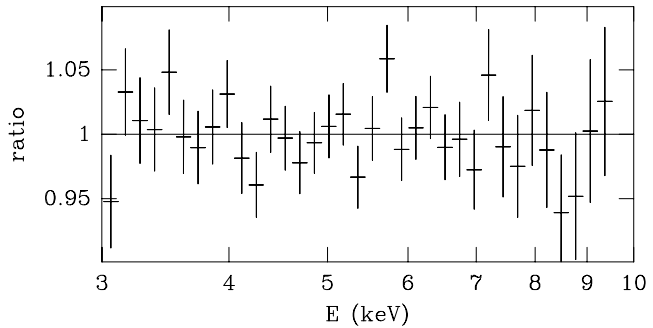}
\end{center}
\caption{The 3--10 keV residuals to the single-line fit above 3 keV to 1993 
May \asca/OSSE data. No broad, redshifted component of the K$\alpha$ line is 
seen. The data from all 4 \asca\/ detectors have been added for plotting. See 
Section 2.3.2. 
 }
\label{fig:res93}
\end{figure} 

On the other hand, Y95 has found that the K$\alpha$ line in this observation is 
very broad and redshifted, which would suggest its origin from Compton 
reflection. As pointed out by Y95, NGC 4151 has a complex continuum shape, and 
the choice of the continuum model affects results regarding the line. In 
particular, the soft X-ray continuum is poorly understood, and our model of it 
is only phenomenological. Thus, we follow Y95 in constraining detailed fits of 
the K$\alpha$ line to the range to $\geq 3$ keV  (but we use the OSSE data as 
a constraint on the continuum). We freeze $\as$, $\is$, and $N_2$ (which 
determine the shape of the spectrum below 3 keV) at the values obtained 
by fitting the whole \asca\/ energy range (Table 1). The single-line fit now 
gives $\chi^2=1158/1231$ d.o.f., and the line parameters are similar to 
those above, $\ef =6.36^{+0.04}_{-0.04}$ keV, $\sf=0.13^{+0.06}_{-0.04}$ keV, 
EW $=220^{+50}_{-50}$ eV. (Other parameters of interest are $\alpha=0.79$, 
$kT=110$ keV, $\af= 1.5\pm 0.3$.) Fig.\ \ref{fig:res93}{\it b} shows the 
resulting 3--10 keV residuals, with no systematic deviations in the 5--6 keV 
range. Indeed, adding a second Gaussian (with 3 free parameters) improves the 
fit by $\Delta\chi^2= -3$ only, which is statistically insignificant. The 
parameters of the second line are $\ef =5.8$ keV, $\sf=0$ keV (poorly 
constrained due to the fact that such a line is not required by the data), and 
EW $=25_{-23}^{+23}$ eV, i.e., the redshifted line component is much weaker 
than the main one. Thus, the presence of a broad and redshifted line component 
with a strength comparable to that of the narrow component is {\it ruled out}. 
This contrasts with the result of Y95, who obtained a fit improvement of 
$\Delta\chi^2=-26$ due to adding a broad and redshifted Gaussian with the 
equivalent width similar to that of the narrow line. 

In order to find the cause of this difference in the results, we have repeated 
the above analysis of the line shape using the same {\sc pha} files and 
response files as those used by Y95. We are still able to obtain a 
satisfactory fit with our single-line model, which gives $\chi^2_\nu= 0.97$ in 
the 3--10 keV range, which is the same as $\chi^2_\nu$ obtained by Y95 for 
their 2-line model. No residuals in the 5--6 keV are seen for that model. 
Adding a second Gaussian results indeed in a negligible fit improvement, 
$\Delta \chi^2= -4$, similarly as in the case of our \asca\/ data.  This shows 
that a major cause of the difference between our results and those of Y95 
is the difference in the choice of continuum. We use a power law (see Fig.\ 
\ref{fig:par} in Section \ref{ss:par}) attenuated by a dual neutral absorber 
with variable Fe abundance whereas Y95 has chosen an analytic function 
designed to reproduce a power law and an ionized absorber. A possible strong 
effect of changing the absorption law on the presence of a broad, redshifted, 
line is illustrated in Fig.\ \ref{fig:res91}. One difference between the 
results based on the two data sets is that the broad, redshifted, Gaussian 
fitted to the data of Y95 has EW $=180$ eV, which is of the same order as the 
EW of the narrow line. Such a strong feature is ruled out in our current 
\asca\/ data. This is explained by the difference between the {\sc ascaarf} 
v.\ 2.53 and 2.61 (see Section \ref{sss:91data}). 
               
However, most of the statistical evidence of Y95 for the presence of a strong 
broad-line component in NGC 4151 comes from 1993 December observation, with 
the exposure time of 41.2 ksec, compared to 15.5 ksec in 1993 May. Thus, our 
conclusions regarding the weakness of a broad line need to be confirmed by 
studying longer-exposure spectra, e.g.\ the 1993 December one. We also stress 
that other Seyferts with a broad line seen in the \asca\/ data, e.g.\ MCG 
--6-30-15 (Tanaka et al.\ 1995), have much less absorption than NGC 4151 and 
thus the shapes of their broad lines are very weakly dependent on details of 
absorption. 

Finally, we note that the data above 3 keV are fitted with $\af=1.5\pm 0.3$, 
which is consistent with $\af=2.2\pm 0.4$ in 1991 June for $\af=1.8$. Thus, 
$\af=1.8$ is the Fe abundance consistent with our 2 sets of data. On the 
other hand, a change of fitted $\af$ may result from our idealized 
modeling of absorption as due to a dual neutral absorber. 

\subsection{Other Data} 
\label{ss:comp} 

The 1991--93 OSSE observations of NGC 4151 show a remarkable constancy of its 
\g-ray flux with time (see Section \ref{sss:osse} below). In the four OSSE 
observations, the 50--200 keV flux changed within $\pm 10$ per cent only. 
This is consistent with the constant intrinsic X\g\ spectrum found in this 
work for the 1991 June and 1993 May observations. On the other hand, \ginga\/ 
has shown strong intrinsic X-ray variability of NGC 4151 (YW91; Y93). Also, 
some previous experiments showed strong \g-ray variability (Perotti et al.\ 
1981, 1991; Baity et al.\ 1984). Those \g-ray data, however, were obtained 
with relatively uncertain background subtraction and low signal-to-noise 
ratio. Thus, a question arises whether the \g-ray flux of NGC 4151 is indeed 
highly variable in general and the present approximate constancy is specific 
to the 1991--1993 epoch or, to the contrary, the \g-ray flux is almost 
constant universally? We attempt here to address this question by considering 
archival OSSE, \ginga, \exosat\/ and \granat\/ data from NGC 4151. 

\subsubsection{OSSE Data}
\label{sss:osse}

All up to-date observations of NGC 4151 by OSSE are presented in detail by 
Johnson et al.\ (1996). Here we consider the data from four viewing periods 
(hereafter VP) until the end of 1993, namely 1991 June 29--July 12 1991 (VP 
4), 1993 April 20--May 3 (VP 218), 1993 May 24--31 (VP 222), and 1993 December 
1--13 [VP 310; from Warwick et al.\ (1996)].  

\begin{table}
\centering
\caption{Fit parameters for the OSSE data for NGC 4151 and the average 
spectrum of weaker radio-quiet Seyfert 1s (Sy1). The model is a power law with 
an exponential cutoff. $\ec$ is the $e$-folding energy in keV, $F_1$ and $F_2$ 
are the 50--100 keV and 100--200 keV fluxes, respectively, in $10^{-4}$ cm$^{-
2}$ s$^{-1}$. 
 }
 \begin{tabular}{lccccc} 
\hline
Obs.  & $\ec$ & $\alpha$ & $\chi^2$/dof
& $F_1$ & $F_2$ \\
  4 
& $83^{+29}_{-22}$ & $0.28^{+0.30}_{-0.41}$
& 42/50& $24$ & $8.7$
\\
218 
& $96^{+80}_{-30}$ & $0.57^{+0.47}_{-0.46}$
& 33/50& $22$ & $7.3$
\\
222 
& $93^{+56}_{-33}$ & $0.30^{+0.42}_{-0.58}$
& 26/50& $25$ & $9.9$
\\
310 
& $70^{+31}_{-22}$ & $0.24^{+0.41}_{-0.61}$
& 38/50& $26$ & $8.5$
\\
All 
& $98^{+18}_{-14}$ & $0.33^{+0.21}_{-0.23}$
& 152/206& --- & ---
\\
Sy1 
& $75^{+12}_{-27}$ & $0.09^{+0.70}_{-0.81}$
& 46/50& --- & ---
\\
Sy1+All 
& $82^{+17}_{-13}$ & $0.31^{+0.20}_{-0.22}$
& 199/258& --- & ---
\\
\hline
\end{tabular}
\label{t:osse}
\end{table}

All the data are well fitted by the thermal Comptonization model with 
reflection. However, we present here fit results with an exponentially cut-off 
power law (which fits the OSSE data alone equally well), in order to give a 
convenient representation of the \g-ray spectra. We stress that this 
phenomenological model should not be extrapolated below 50 keV, and that 
$\alpha$ cannot be interpreted as the X-ray spectral index implied by the 
data. Table 2 shows the fit results as well as the photon fluxes in 50--100 
keV and 100--200 keV bands. We see that all the spectra are approximately 
compatible with the constant shape, and the fluxes vary within $\pm 10$ per 
cent only. All four data sets can indeed be well fitted ($\chi_\nu^2= 0.73$) 
with the same model at free normalization, as shown in Table 2. 

Table 2 also shows the fit results for the average spectrum of all radio-quiet 
Seyfert 1s observed so far by OSSE with the exception of NGC 4151 
(McNaron-Brown et al.\ 1996). That spectrum has been obtained using the current 
OSSE response. The response revision has resulted in an overall spectral 
softening with respect to the average Seyfert spectrum presented by Johnson et 
al.\ (1994). We see that the spectral parameters are consistent with those of 
NGC 4151 within the statistical uncertainties, as pointed out by Johnson et 
al.\ (1994). This can be quantified by fitting simultaneously the average 
Seyfert-1 spectrum together with all 4 spectra of NGC 4151. We find that the 
fit is then almost identical to that for the four NGC 4151 spectra at $\Delta 
\chi^2<1$ with respect to the $\chi^2$ sum of the previous two individual 
fits. Thus, the OSSE data do not show a difference between the spectra of NGC 
4151 and that of of the average Seyfert 1 sample. We have also obtained the 
same results for the average radio-quiet Seyfert-1 spectra based on smaller 
samples of objects observed by both \ginga\/ and OSSE and by \exosat\/ and 
OSSE (5 and 7 objects, respectively; Gondek et al.\ 1996). 

\subsubsection{Ginga and Exosat data}
\label{ss:exg}

The 2 keV flux in \exosat\/ and \ginga\/ data changes by a factor of $\sim 35$ 
(Fig.\ \ref{fig:exg}; Pounds et al.\ 1986; Yaqoob, Warwick, \& Pounds 1989; 
YW91; Y93). During the 1991--1993 monitoring of NGC 4151 by OSSE, four X-ray 
observations (Y93; W94; Y95; Warwick et al.\ 1996) show the 2 keV flux varying 
within a factor of $\sim 5$ in a lower part of the flux range observed by 
\exosat\/ and \ginga. 

\begin{figure*}
\begin{center}
\leavevmode
\epsfxsize=9.5cm \epsfbox{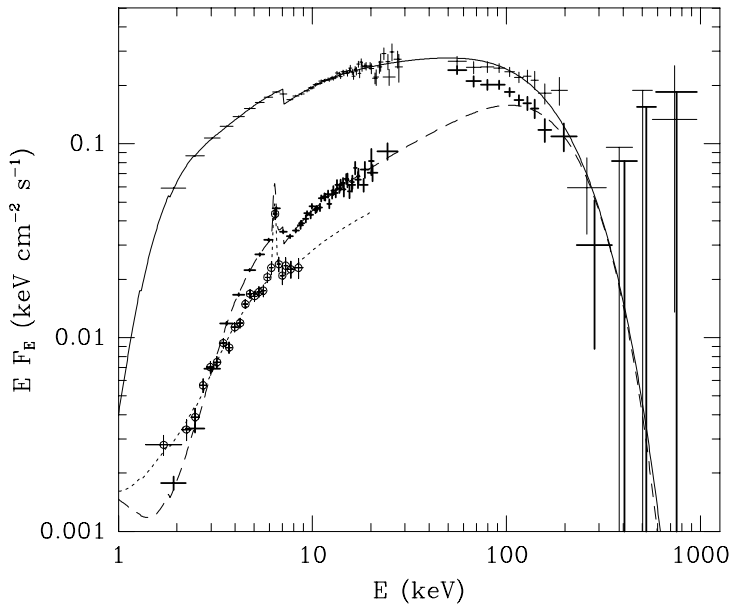}
\end{center}
\caption{The range of X-ray states of NGC 4151 seen by \ginga\/ (thin and 
thick crosses at $<30$ keV) and \exosat\/ (circled crosses) together with 
the range of \g-ray states seen by OSSE (not simultaneous; thin and thick 
symbols above 50 keV). The solid curve represents the thermal Comptonization 
model, which fits well the combined highest X\g\ data. The dashed and dotted 
curves represent fits of the same model to the lowest \ginga\/ and \exosat\/ 
data, respectively, which both imply $\sim 50$ keV fluxes below the range seen 
by OSSE. See text. 
 }
\label{fig:exg}
\end{figure*} 

If the \g-ray flux were universally close to constant, one should be able to 
fit the archival \ginga\/ data together with the current OSSE data. We use 
both top-layer and mid-layer \ginga\/ data. We first consider the data with 
the highest 2--10 keV X-ray flux, which were obtained 1990 May 15--16 (Y93). 
Fig.\ \ref{fig:exg} shows the fit of the thermal Comptonization model. For 
models in this section, the soft excess component is fixed at the shape 
obtained for 1991 June data (see Table 1), and the column $N_2$ is constrained 
to $\geq 10^{22}$ cm$^{-2}$ to avoid the hard X-ray continuum to appear below 
$\sim 1$ keV (see YW91). We use the OSSE 1993 May 24--31 data, which have the 
highest 50--200 keV flux. We obtain $\alpha= 0.84^{+0.04}_{-0.03}$ and $kT= 
78^{+48}_{-17}$ keV at $\chi^2= 84/104$ d.o.f., i.e., the data are very well 
fitted by the model. Thus, the highest observed X-ray and \g-ray states are 
compatible with each other. There is no reflection ($R=0^{+0.18}$ at 
$i=65^\circ$ and $R=0^{+0.08}$ at $i=20^\circ$), which is compatible with the 
lack of a detectable Fe K$\alpha$ line (EW $=0^{+15}$ eV). (Fitting the 
\ginga\/ data alone gives almost identical results for $\alpha$ and $R$.) The 
upper limit on a second broad and redshifted line (see Section \ref{ss:93}) 
corresponds to EW $=0^{+16}$ eV. The covering factor is about 0.3, which is 
much less than typical for NGC 4151 (Y93). The Fe abundance is $4.3^{+0.9}_{-
1.0}$, which is not compatible with the range of $\af\sim 1$--2.6 found for 
other data considered here (see also Y93). Since we use this data set only to 
constrain the variability range, we do not attempt here to explain this 
peculiar behaviour. 

Then we fit the \ginga\/ data with the lowest X-ray flux, which was observed 
1987 May 29--31 (YW91). The \ginga\/ data are fitted with $\alpha= 
0.54^{+0.15}_{-0.06}$ (for the assumed $kT=60$ keV), $R=0^{+0.30}$ for 
$i=65^\circ$ ($R^{+0.15}$ for $i=20^\circ$), and $\af=2.1^{+0.5}_{-0.5}$ at 
$\chi^2=48/49$ d.o.f. There is a strong K$\alpha$ line with EW $= 270_{-
80}^{+60}$ eV. The spectrum is strongly absorbed, with the covering factor of 
about 0.8. When extrapolated above 30 keV, the \ginga\/ data imply 
$EF_E(50\,{\rm keV})= 0.12\pm 0.02$ keV cm$^{-2}$ s$^{-1}$, which is a factor 
of 2 below the lowest 50 keV flux in the OSSE data up to 1993 (1993 April 
20--May 3), $EF_E=0.24\pm 0.01$ keV cm$^{-2}$ s$^{-1}$, as shown in Fig.\ 
\ref{fig:exg}.  

Interestingly, the two \ginga\/ observations do not confirm the presence of 
continuum Compton reflection in NGC 4151. Furthermore, the flux in the 
K$\alpha$ line varies less than the X-ray continuum, as well as it is larger 
for more absorbed spectra. This is inconsistent with the origin of most the 
line from reflection but rather suggestive of a substantial contribution from 
absorption by cold matter at a relatively large distance from the nucleus. 
Some K$\alpha$ emission is then due to absorption of the continuum, and it is 
(i)  averaged over the light-travel time across the absorber and (ii) stronger 
when absorption is stronger (Makishima 1986). 

All X-ray spectra obtained by \exosat\/ are below the spectrum of the highest 
\ginga\/ state. On the other hand, the lowest X-ray states from \exosat\/ 
obtained in 1984 April (Pounds et al.\ 1986) have 2--10 keV fluxes lower than 
that of the lowest \ginga\/ state. There are two data sets, from April 8 and 
April 18--19. The spectrum with the lower flux (April 18--19) is indicated as 
uncertain (quality flag 2) in the \exosat\/ archive. Thus, we use here the 
spectrum from April 8 (quality flag 4), which has the 2--10 keV flux only 10 
per cent higher (although both spectra are similar). The data quality is, 
however, still insufficient to constrain the Fe abundance. Thus, we fix it at 
$\af=2$ (see Table 1; Y93). Then a fit with a power law, dual absorber and the 
fixed soft excess yields $\alpha=0.47_{-0.33}^{+0.57}$, and a strong K$\alpha$ 
line with EW $=350$ eV, see Fig.\ \ref{fig:exg}. Although the implied 50 keV 
flux is poorly constrained due to the uncertain $\alpha$, even the hardest 
power law allowed by the data yields $EF_E(50\,{\rm keV})= 0.14$ keV cm$^{-2}$ 
s$^{-1}$, which is a factor of $\sim 2$ below the lowest 1991--93 OSSE state, 
similarly to the lowest \ginga\/ state. 

\subsubsection{GRANAT data}
\label{ss:granat}

NGC 4151 was observed from 1990 July to 1992 November by the \granat\/ 
satellite (Finoguenov et al.\ 1995). In particular, it was observed by the 
ART-P (below 30 keV) and SIGMA (above 30 keV) instruments on 1991 June 29 and 
July 11--12 , i.e.,  contemporaneously with the OSSE observation of 1991 June 
29--July 12. The \granat, OSSE, and \ginga\/ observations are compared in 
Fig.\ \ref{fig:granat}. We see that the OSSE and SIGMA data are consistent 
with each other. On the other hand, the ART-P data are somewhat above the 
corresponding \ginga\/ data. The cause of that may be X-ray variability (since 
the \ginga\/ data are from about a month earlier) as well as uncertainty in 
the relative calibration of the instruments. Still, the \ginga\/ data match 
very well the OSSE data (see Fig.\ \ref{fig:spec}{\it a}), and the slow 
variability seen in the OSSE range (Section \ref{sss:osse}) is compatible with 
the spectrum above 50 keV during the \ginga\/ observation being indeed at the 
level measured one month later. 

SIGMA has observed an almost constant \g-ray flux from NGC 4151 during 4 out 
of 5 observations (Finoguenov et al.\ 1995). During those observations, $EF_E$ 
at 100 keV was $\sim 0.16\pm 0.05$ keV cm$^{-2}$ s$^{-1}$. In 1991 November, 
the flux of $0.30\pm 0.04$ keV cm$^{-2}$ s$^{-1}$ was observed. The 
corresponding range in the 1991--93 OSSE observations is $0.18\pm 0.01$
to $0.25\pm 0.01$ keV cm$^{-2}$ s$^{-1}$. Thus, although the SIGMA data by 
themselves indicate stronger variability than the OSSE data, the two data sets 
are consistent with each other within measurement errors. 

Summarizing, the \exosat, \ginga\/ and \granat\/ results strongly suggest that 
the fluxes between 50 and 100 keV vary at least within a factor of about 2, 
which is more than the range observed by OSSE in 1991--93. The strongest 
prediction is that of $EF_E(50\,{\rm keV})\simeq 0.12\pm 0.02$ keV cm$^{-2}$ 
s$^{-1}$ from extrapolation of the lowest \ginga\/ spectrum, which $EF_E$ is 
factor of 2 below the lowest 1991--93 OSSE flux of $0.24\pm 0.01$ keV cm$^{-
2}$ s$^{-1}$. This is fact is confirmed by OSSE observation of a low $>50$ keV 
state in 1995 (Johnson et al.\ 1996).

\begin{figure}
\begin{center}
\leavevmode
\epsfxsize=8.4cm \epsfbox{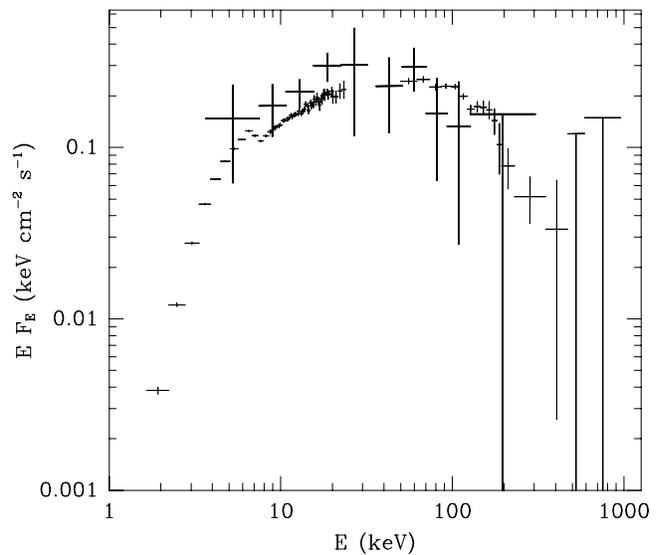}
\end{center}
\caption{Comparison of observations of NGC 4151 by \granat\/ 1991 June 29 and 
July 11--12 (thick symbols), by OSSE June 29--July 12 (thin symbols above 50 
keV), and by \ginga\/ May 31-June 1 (thin crosses below $30$ keV). 
 }
\label{fig:granat}
\end{figure}

\section{THEORETICAL IMPLICATIONS}
\label{s:theory}

\subsection{Source parameters and geometry}
\label{ss:par}

Our thermal Comptonization model uses $\alpha$ and $kT$ as free parameters. 
The plasma optical depth, $\tau$, is geometry-dependent. To determine it 
accurately for a uniform spherical source, we use a Monte Carlo method (see 
Appendix). We will use throughout the continuum parameters derived for 1991 
June as the parameters for 1993 May are very similar, see Table 1. 

We find that the spectrum observed in 1991 June (with $\alpha=0.80$ and $kT= 
88$ keV) corresponds to that from Componization in a uniform sphere (with a 
uniform distribution of seed-photon sources) with $kT= 61$ keV and $\tau=1.3$. 
The temperature is lower than that obtained from fittting because the 
Comptonization model is formally obtained under the assumption of $\tau^2\gg 
1$ whereas $\tau$ approaches unity now. Still, that model gives a very good 
description of the shape of the spectrum, albeit for a somewhat different 
temperature, as shown in Fig.\ \ref{fig:par}. Thus, the fit results for $kT$ 
in Table 1 need to be rescaled by about 2/3, similarly to the case of the 
approximation of emission from optically thin plasmas by a power law with an 
exponential cutoff (Zdziarski et al.\ 1994). (However, a power law with an 
exponential cutoff does {\it not\/} approximate emission from plasmas with 
$\tau\simgreat 1$.) The agreement of the parameters of our model with those 
from Monte Carlo improves rapidly with increasing $\tau$ (see Appendix). 

\begin{figure}
\begin{center}
\leavevmode
\epsfxsize=8.4cm \epsfbox{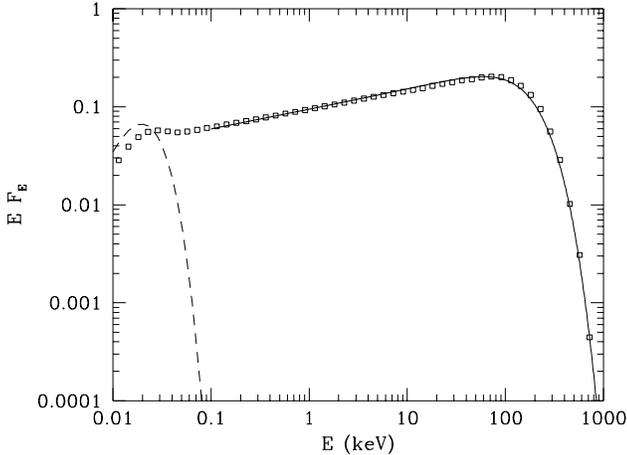}
\end{center}
\caption{The Comptonization spectrum from a spherical plasma cloud with 
parameters corresponding to 1991 June. The solid curve gives the spectrum of 
the fitted continuum models and squares give results of our Monte Carlo 
simulations, see text. The model spectrum agree very well with Monte Carlo 
simulations. The dashed curve represents the seed diluted-blackbody photons 
with the temperature of 5 eV.
 }
\label{fig:par}
 \end{figure} 

The plasma cloud is irradiated by soft photons providing seeds for 
Comptonization. We take the seed photons as diluted blackbody at the 
temperature of 5 eV (Zdziarski \& Magdziarz 1996; Kriss et al.\ 1995). The 
ratio of the Comptonized flux to the the soft flux irradiating the plasma is 
13, i.e., the plasma is soft-photon starved. Some or all of the seed 
photons are due to reprocessing of the Comptonized, hard, radiation by cold 
matter, e.g.\ an accretion disk. We consider this effect here and find that 
the X\g\ source cannot, contrary to a claim of TM94, form a homogeneous corona 
above the surface of an accretion disk. 

A hot corona above a cold disk is cooled by soft photons emitted by the disk. 
The soft emission is due to both internal dissipation in the disk and 
reprocessing of hard radiation from the corona irradiating the disk. The 
cooling is minimized when all the dissipation occurs in the corona and none in 
the disk (Haardt \& Maraschi 1993), which case we consider below. Anisotropy 
effects (Haardt 1993) are small at our derived plasma parameters and the 
corona emits photons approximately equally up and down (towards the disk). The 
down part is mostly absorbed and reemitted as the seed thermal UV radiation. 
The seed photons are upscattered, which self-consistently forms the X\g\ 
corona emission. Since the seed photons are from the reprocessed $\sim 50$ per 
cent fraction of the corona photons emitted down, the observed X\g\ 
luminosity, $\lxg$ (from the $\sim 50$ per cent of the corona photons emitted 
up), equals approximately the UV disk luminosity, $\luv$. 

This condition is strongly violated in NGC 4151, in which $\luv\ll \lxg$ (see 
above). We have performed detailed Monte-Carlo calculations of a slab of hot 
electrons above a cold disk. For the spectral parameters for the 1993 June 
observation, we find the full slab thickness corresponds to $\tau\simeq 1.3$. 
The slab luminosity directed downward is 8.0 times the seed UV luminosity, 
$\luv^{\rm seed}$. Then we compute the integrated albedo of the cold disk, 
$A=0.22$. Thus, the UV emission of the cold disk from reprocessing of the 
corona hard emission is $8(1-0.22)\simeq 6.2$ times the seed UV luminosity, 
i.e., $\luv^{\rm reprocessed}\simeq 6 \luv^{\rm seed}$. This strongly violates 
the energy balance condition, $\luv^{\rm seed}= \luv^{\rm reprocessed}$. The 
above discrepancy in the luminosities is much worse for the spectrum used by 
TM94, which is much harder than our intrinsic spectrum, and thus much more 
soft-photon starved. Furthermore, any power internally dissipated in the disk 
will worsen this discrepancy. Thus, the high-energy source in NGC 4151 cannot 
form a homogeneous corona, contrary to TM94. The discrepancy is due to TM94 
both neglecting bound-free absorption in their albedo (which underestimates 
the absorbed fraction, $1-A$) and using formulae of Titarchuk (1994) derived 
under the assumption of $1-A\ll 1$ (while the actual $1-A\simeq 0.8$). (We 
also note that $\tau=1.25$ given in TM94 as the optical depth of their hot 
corona is in fact the half-thickness of the corona on each side of the disk.) 

Thus, the spectrum of NGC 4151 requires that the reprocessed UV radiation {\it 
returning\/} to the hot source(s) is reduced to $\sim 0.2$ with respect to 
that from a hot slab located above a cold slab. This soft-photon starvation 
can be achieved by geometry. The source can form one or more cloud at some
height above the disk (e.g., Svensson 1996). For a suitable ratio of the 
height to the source size, only $\sim 20$ per cent of the reprocessed 
emission returns to the hot source. [Note that this geometry is different from 
the patchy coronae of Haardt, Maraschi \& Ghisellini (1994) and Stern et al.\ 
(1995), in which X\g\ sources are close to the disk surface and most of the 
reprocessed emission does return to the hot source.] On the other hand, the 
disk-corona geometry would imply a Compton reflection component with $R\sim 
1$, which is more than that observed (unless for an edge-on orientation; 
Sections \ref{ss:91}, \ref{ss:exg}). If the cold medium is Thomson-thick, the 
weakness of Compton reflection implies that the cold medium covers at most a 
$\sim 1\pi$ solid angle as seen from the hot source. A geometry compatible 
with both soft photon-starvation and reduced reflection is a hot inner disk 
and a cold outer disk (e.g., Shapiro, Lightman \& Eardley 1976). 
Alternatively, the covering by the cold medium could be higher if the medium 
is Thomson-thin. Also, the hardness of the X-ray spectrum together with the 
relative weakness of Compton reflection is compatible with some configuration 
of hot and cold clouds with the mutual covering factors less than unity. 

\subsection{Electron-positron pairs}
\label{ss:pair}

Here we examine if the hot source in NGC 4151 is dominated by \ee\ pairs or 
by electrons. The pair production rate depends on the shape and the 
amplitude of the spectrum (Gould \& Schr\'eder 1967). On the other hand, the 
pair annihilation rate depends on the optical depth of the source. The two 
quantities are equal in pair equilibrium, which is established under most 
conditions even in variable sources (Svensson 1984). 

We first assume that the X\g\ emission is due to purely thermal 
Comptonization. We use the Comptonized spectrum with the parameters as for 
1991 June (Table 1) to compute the total pair production rate (Gould \& 
Schr\'eder 1967). This rate scales with the size of the X\g\ source, $\rxg$, 
as $\rxg^{-1}$. On the other hand, the pair annihilation rate scales as 
$\rxg$. Requiring the pair equilibrium we can solve for $\rxg$. Assuming a 
spherical pure pair source and setting the average photon escape time to 
$\rxg/c$ we obtain a value of $\rxg=2.2\times 10^{11}$ cm. The X\g\ luminosity 
is $\lxg \simeq 5\times 10^{43}$ erg s$^{-1}$, and the corresponding 
compactness parameter is very large, $\ell \sim 10^4$, where $\ell \equiv \lxg 
\sigma_{\rm T}/(\rxg m_e c^3)$ (Svensson 1984). Thus, the required size of the 
pair source is rather small, and much less than the characteristic size of 10 
Schwarzschild radii, $10\rs\sim 10^{14}$ cm implied by the black hole mass 
estimate of $\sim 4\times 10^7 M_\odot$ of Clavel et al.\ (1987). Even the 
size inferred from the minimum possible mass from the condition of 
sub-Eddington luminosity during the highest states of the source (Section 
\ref{ss:comp}), $10\rs\sim 10^{13}$ cm, is still much more than the required 
size of the pair source. Furthermore, the Eddington limit is reduced for 
pair-dominated sources (Lightman, Zdziarski \& Rees 1987), in which case $10 
\rs \gg 10^{13}$ cm. From the causality arguments it seems unlikely to release 
most of the bolometric luminosity (which is mostly in X\g\ photons) in a 
region much smaller than the region in which most of the gravitational energy 
is released. 

Thus, if the X\g\ source in NGC 4151 is thermal, it consists most likely of 
e$^-$ rather than \ee\ pairs. Note that the situation in NGC 4151 differs 
strongly from that assumed in the models of Seyfert 1s of Stern et al.\ 
(1995). That study considers plasmas hotter and optically-thinner than that 
found here for NGC 4151, which results in $\ell\sim 10$ being already 
sufficient for the dominance of pairs. 

On the other hand, ZLM93 considered a hybrid, thermal/nonthermal, model for NGC 
4151, in which a small fraction of the electrons or \ee\ pairs in the source 
is accelerated to nonthermal, relativistic, energies, but the bulk of the 
spectrum is still produced by thermal Comptonization. In the limit of zero 
nonthermal fraction (i.e., uniform heating of all electrons in the source), 
the model yields a pure thermal Comptonization spectrum. Acceleration has a 
two-fold role in the model. First, Compton scattering by nonthermal electrons 
as well as \ee\ pair annihilation give rise to a weak tail on top of the 
thermal-Comptonization spectrum. [This contrasts the pure-nonthermal pair model 
proposed earlier to explain the \ginga\/ spectra of typical Seyfert 1s 
(Zdziarski et al.\ 1990), in which {\it most\/} of the X\g\ spectrum is due to 
nonthermal Comptonization.] Second, pair production by \g-rays in the tail can 
supply some or all thermal electrons/pairs in the source. 

ZLM93 found that addition of a tail from nonthermal acceleration improves the 
fit to the \ginga/OSSE spectrum of 1991 June above $\sim 200$ keV. However, 
the plasma temperature in our present model (Table 1) is much above $kT \sim 
40$ keV obtained by ZLM93. The present higher value is due to inclusion of 
Compton reflection, the revision of the OSSE response, as well as using a 
better absorber model. Consequently, our present model spectra (see Fig.\ 
\ref{fig:spec}) have the high-energy cutoff much more gradual than that of the 
spectrum of ZLM93. Thus, a spectral tail in our present model is not required. 

However, an attractive feature of the hybrid model is that it can provide the 
electron optical depth needed for thermal Comptonization self-consistently. We 
find that the nonthermal compactness, $\lnt \equiv L_{\rm nth} \sigma_{\rm 
T}/(\rxg m_e c^3)$ (where $L_{\rm nth}$ is the the nonthermal luminosity), 
required to produce pairs with the needed optical depth is $\lnt\simeq 10$, 
approximately independently of the total compactness. The nonthermal fraction, 
$\lnt/\ell$, is constrained to be small by the form of the cutoff in the 
\g-ray spectrum. For the \ginga/OSSE spectrum, we obtain $\lnt/\ell 
=0.08^{+0.11}_{-0.08}$ in a model with nonthermal electrons accelerated to the 
Lorentz factor of $10^3$. The best fit corresponds to a pure pair source with 
$\ell \simeq 140$ and $\Delta\chi^2= -0.5$ with respect to the pure-thermal 
model, and the lower limit corresponds to the pure thermal-e$^-$ source. 

Thus, the source can be dominated by \ee\ pairs if just $\sim 10$ per cent of 
the available power is released nonthermally. Such a situation is likely if 
the X\g\ emission is from magnetic flares, in which reconnection can 
accelerate electrons. The characteristic compactness is $\ell\sim 10^2$, which 
corresponds to $\rxg\sim 2\times 10^{13}$ cm, which is then compatible with 
the estimates of the source size above. The compactness of $\ell\sim 100$ is 
super-Eddington for a {\it pure\/} pair source and thus we can rule out such a 
case. However, just $\sim 10$ per cent of protons in a pair-dominated source 
can provide the needed gravitational confinement of the plasma at that 
compactness (Lightman et al.\ 1987). Thus, the hybrid model described above is 
still a viable alternative to the pure thermal-e$^-$ model. 
                    
\section{DISCUSSION AND CONCLUSIONS}
\label{s:dis} 

Observations discussed in this paper fall into two groups. First, we have 
found that the intrinsic spectra of NGC 4151 in 1991 June and 1993 May are 
the same within the observational uncertainties. This intrinsic spectrum is 
rather typical for Seyfert 1s (Nandra \& Pounds 1994; Zdziarski et al.\ 1995) 
with an X-ray spectral index of $\alpha\simeq 0.8$, a Compton-reflection 
component, and the soft \g-ray spectrum with the shape not distinguishable 
from other Seyfert 1s. This is a new result for NGC 4151. On the other hand, 
archival data do show the intrinsic X-ray spectrum to be strongly variable, 
$\alpha\sim 0.3$--0.8, and very hard in low states, as well as with no 
reflection component at least in some states (confirming earlier results, 
e.g., YW91, Y93). 

The Compton-reflection component found in the 1991 June \ginga/OSSE spectrum 
is relatively weak. It can be due to reflection from a disk (covering a solid 
angle close to $2\pi$) if the nucleus of NGC 4151 is oriented close to 
edge-on. For a face-on orientation, the covering factor of the reflecting 
medium is only $\sim 0.2\times 2\pi$. The fitted reflection fraction cannot be 
increased by varying either the Fe abundance, the overall metal abundance, or 
the ionization state of the reflector (Section \ref{ss:91}). The shape of the 
continuum does not allow a determination of the inclination angle.  

In spite of the similar continua in the 1991 and 1993 spectra, the Fe K$\alpha$ 
line is much stronger in the 1993 observation, in which X-ray absorption is 
also stronger. The line in 1993 is narrow, and any broad and redshifted 
component (not required by the data) is constrained to $\simless 20$ per cent 
of the narrow component. The narrowness of the line in the 1993 observation 
together with the lack of the correlation between the strength of the 
continuum and that of the line suggests the origin of most the line photons
from absorption rather than reflection. This is consistent with the relatively 
weak reflection in the 1991 June data. 

The X\g\ continuum is very well modeled by thermal Comptonization of soft 
UV photons by a plasma with $\tau\sim 1$ and $kT\sim 60$ keV (Section 
\ref{s:theory}) and attenuated in X-rays by a dual neutral absorber (which 
approximates absorption by a distribution of clouds). We have considered 
alternative models with an exponentially cut-off power law instead of the 
Comptonization spectrum as well as with an ionized absorber instead of the 
dual neutral absorber. We have found those alternative models describe the 
data much worse. 

Another alternative model of the X-ray spectra in 1991 June and 1993 May has 
been proposed by Poutanen et al.\ (1996, hereafter P96). They assumed the 
intrinsic X\g\ emission to be completely attenuated by an optically 
thick medium, and the observed emission to be from scattering towards our line 
of sight by an optically thin plasma, similarly as in the unified model of 
Seyfert 2s (e.g.\ Antonucci 1993). The shape of the observed spectrum is that 
of the intrinsic one at low energies, but it has an additional cutoff due to 
the Klein-Nishina scattering at high energies (e.g., Jourdain \& Roques 1995). 
P96 consider two intrinsic emission models. One is a power law 
($\alpha=0.67^{+0.07}_{-0.05}$, which is harder than the Seyfert-1 average) 
with an exponential cutoff and reflection (as well as a soft X-ray component 
and a K$\alpha$ line), and the other is a self-consistent disk-corona emission 
model (with $kT\simeq 200$ keV). Both models have high-energy cutoffs at 
several hundred of keV, which corresponds to the  average spectrum, with 
$\alpha \simeq 0.9$, of Seyfert 1s (Zdziarski et al.\ 1995; Gondek et al.\ 
1996). The models fitted to the \rosat/\ginga/OSSE data of 1991 June (but 
without the \ginga\/ mid-layer data) yield $\chi^2=120/110$ d.o.f.\ and 
$124/111$ d.o.f., respectively. Those fits are much worse that these obtained 
by us (Section \ref{sss:91res}). Since P96 use a different model for the soft 
X-ray component and do not show their residuals or the contributions to 
$\chi^2$ from each detector, it is difficult to establish the cause of the 
large difference in $\chi^2$. However, the self-consistent disk-corona model 
plotted in P96 clearly overpredicts the spectrum around 20 keV, and it appears 
to give a worse fit to the OSSE data than our model does. P96 also fit their 
model to the \asca/OSSE data of 1993 May. However, they use the SIS0 data with 
their own normalization, which is known to be incorrect (W94). That 
normalization is, in fact, a factor of $\sim 2$ less than that of the correct 
spectrum (Fig.\ \ref{fig:spec}{\it b}), and thus we do not compare our results 
for 1993 May with those of P96. 

The difficulties of the scattering, Seyfert--2-like, model in fitting NGC 4151
appear to reinforce our conclusion that NGC 4151 as observed in 1991 and 1993 
is intrinsically a Seyfert 1 with an average X\g\ spectrum. The spectral index 
of $\alpha \simeq 0.8$ is within the 1-$\sigma$ range of Seyfert 1s, $0.95\pm 
0.15$ (Nandra \& Pounds 1994). Also, the OSSE spectra of NGC 4151 are 
statistically the same as the average spectrum of weaker radio-quiet Seyfert 
1s observed by OSSE. The fact that the OSSE spectra of NGC 4151 are fitted 
with a high-energy cutoff energy less than that of the average Seyfert-1 X\g\ 
spectrum is explained by a correlation between $\alpha$ and the cutoff 
energy. The correlation appears because harder X-ray spectra (such as of NGC 
4151) need to be cut off faster than softer X-ray spectra in order to fit the 
same \g-ray spectrum. 

The X-ray spectrum of NGC 4151 is too hard for the homogeneous disk-corona 
model of Haardt \& Maraschi (1993) to apply. (We find that the disk-corona 
model applied to NGC 4151 by TM94 is in error.) We find the X\g\ source in NGC 
4151 subtends a small solid angle of $\sim 0.2 \times 2\pi$ as seen from the 
UV source (providing seed photons for Comptonization). This also rules out 
patchy corona models with the X\g\ sources located on the surface of the disk. 
The possible geometries are either the X\g\ sources at a large enough height 
above the surface of an accretion disk (e.g., Svensson 1996), a hot inner 
disk, or the UV-emitting cold matter in form of clouds. 

Similar geometries have been considered by Zdziarski \& Magdziarz (1996) as 
models for the UV/X-ray correlation observed in NGC 4151 by \exosat\/ and 
\iue\/ (Perola et al.\ 1986). Their transmission model explains the 
correlation as being due to the X-ray absorber re-emitting the absorbed X-ray 
power in the UV. The source parameters obtained here for the 1991 June and 
1993 May data are consistent with that model. In particular, the width of the 
K$\alpha$ line implies its origin at $\simgreat 100$ Schwarzschild radii 
(Section \ref{sss:93res}), which is compatible with the UV response to varying 
X-rays delayed by $\simless 0.3$ day (Edelson et al.\ 1996) for the black hole 
mass of $\sim 10^7 M_\odot$. Unfortunately, no UV data exist for the X\g\ 
observations studied here. The reflection model of Zdziarski \& Magdziarz 
(1996) explains the correlation as UV-reemission of X\g\ emitted towards a 
cold disk by a patchy corona. That model is independent of the corona 
geometry, and thus it is compatible with a photon-starved corona required by 
the spectra reported here. The model also requires that $R\cos i\simgreat 
0.15$, which condition is satisfied for the 1991 June observation (Section 
\ref{sss:91res}). We intend to test the models of Zdziarski \& Magdziarz 
(1996) against the 1993 December \iue/\rosat/\asca/OSSE data (Warwick et al.\ 
1996; Edelson et al.\ 1996). 

The parameters of the X\g\ source ($\tau\sim 1$, $kT\sim 60$ keV, Section 
\ref{ss:par}) imply that pair production is negligible if the plasma is 
thermal (Section \ref{ss:pair}). On the other hand, a small fraction of the 
total power used to accelerate electrons to relativistic energies results in 
enough pair production for the plasma to be pair-dominated. Such acceleration 
may take place in magnetic flares, which are likely to be responsible 
for the formation of the disk corona. In both cases, the process responsible 
for the continuum emission is {\it thermal\/} Comptonization. 

The conclusion of negligible thermal pair production differs from that of 
Stern et al.\ (1995), whose models are pair-dominated at $\alpha=0.8$. 
Their models have $\tau\ll 1$ and $kT\sim 511$ keV, which allows them to be
pair-dominated at a relatively low compactness. These models can fit the 
average OSSE spectrum of radio-quiet Seyfert 1s of Zdziarski et al.\ (1995) 
and Gondek et al.\ (1996). However, we are able to rule out $\tau\ll 1$ by 
direct fitting to the NGC 4151 spectrum, which is of much higher statistical 
accuracy. 

We have also compared our 1991 and 1993 multiwavelength observations with 
archival \ginga\/ and \exosat, \granat\/ and OSSE data. It appears that the 
intrinsic spectrum with $\alpha\simeq 0.8$ of 1991 and 1993 belongs to a high, 
soft, state of NGC 4151. The archival data also show a continuum of states 
limited by a low, hard, state with $\alpha\simeq 0.3$ (as found before by 
YW91 and Y95). The hardening of the X-ray spectrum with the decreasing X-ray 
flux results in the spectrum pivotting at a few hundred keV (YW91; Y93). 
This property allows the \g-ray spectrum, which breaks at $\sim 100$ keV, to 
vary within to a factor of a few only (as observed by OSSE and \granat), 
while the X-rays vary much more strongly (see Fig.\ \ref{fig:exg}). This 
behaviour can be modeled, for example, as due to a strongly-varying soft (UV) 
photon flux irradiating the X\g\ source with a weakly-varying X\g\ luminosity. 

The hardness of the X-ray spectrum observed in some states of NGC 4151 by 
\ginga\/ and \exosat\/ is not unique among AGNs, and it is similar to that of, 
e.g., some Seyfert 2s (Smith \& Done 1996). An interesting example is MCG 
--5-23-16. Two X-ray spectra from \ginga\/ (Smith \& Done 1996) are compared 
with the OSSE spectrum (Johnson et al.\ 1994) of that AGN in Fig.\ 
\ref{fig:52316}. The spectral indices are 0.44 and 0.62 (for a fit with a 
power law with an exponential cutoff) for the lower and higher X-ray state, 
respectively. This is within the range observed in NGC 4151 (Section 
\ref{ss:exg}). We see that the variable X-ray power law pivots around 100 keV. 
As a consequence, both \ginga\/ spectra can be fitted together with the OSSE 
spectrum (not simultaneous). Also, the K$\alpha$ line in MCG --5-23-16 has the 
constant line flux in the two different continuum states, as well as there is 
no detectable Compton reflection (Smith \& Done 1996), which is similar to 
some states of NGC 4151 (Section \ref{ss:exg}). 

\begin{figure}
\begin{center}
\leavevmode
\epsfxsize=8.4cm \epsfbox{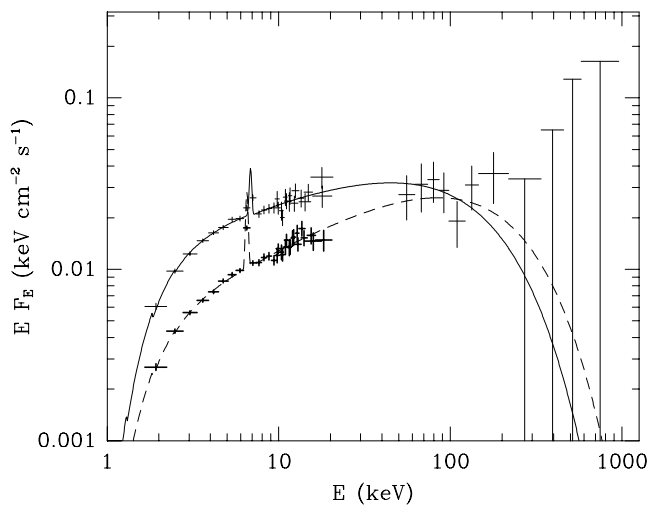}
\end{center}
\caption{Spectra of MCG --5-23-16 (a Seyfert 2) from two \ginga\/ and one OSSE 
observations (not simultaneous). The dashed and solid curves represent the 
fits with $\alpha\sim 0.4$ and 0.6, respectively. Such spectra are within the 
range seen in NGC 4151, see Fig.\ 4. 
 }
\label{fig:52316}
 \end{figure}

\section*{ACKNOWLEDGMENTS}
This research has been supported in part by NASA grants and contracts and the 
Polish KBN grants 2P03D01008 and 2P03D01410. It has made use of data obtained 
through the High Energy Astrophysics Science Archive Research Center Online 
Service, provided by NASA/GSFC. We are especially grateful to C. Done, who 
provided very valuable comments and suggestions pertaining to this work, as 
well as supplied us with the \rosat\/ data. Also thanks are due to D. Smith, 
who provided us with the \ginga\/ data, T. Yaqoob, who supplied us with 
his \asca\/ data files (used in Y95), K. Ebisawa for help with {\sc ascaarf}, 
and Greg Madejski for help with the \asca\/ data reduction.

\section*{APPENDIX: OPTICALLY THICK THERMAL COMPTONIZATION}

We use the thermal Comptonization model of LZ87 to fit the spectra of NGC 
4151. That work uses the Kompaneets equation with a relativistic correction to 
energy transfer between photons and electrons [eq.\ (22) in LZ87)]. In the 
relation between the photon density inside the source and the photon escape 
rate, LZ87 also take into account reduction at relativistic photon energies of 
the photon build-up inside the source due to diffusion [eq.\ (21) in LZ87]. 
The Kompaneets equation is then solved numerically for an assumed distribution 
of seed photons, $kT$, and $\tau$. 

LZ87 use a photon-escape probability formalism to obtain the spectra of 
escaping photons. This implies that the $\tau$ used in the Kompaneets equation 
corresponds to a source with the assumed escape probability. Thus, that $\tau$ 
is geometry-dependent, and it approximately equals the radial optical depth 
in a uniform sphere. A property of the Kompaneets equation that is 
geometry-independent is the spectral index of the asymptotic low-energy photon 
spectrum, $\alpha$ (at energies above those of the seed photons). In the 
model of LZ87, it equals, 
 \begin{equation}
\alpha=\left[ {9\over 4} +{1\over (kT/m_e c^2)\tau(1+\tau/3)} \right]^{1/2} - 
{3\over 2}
\label{alpha}
\end{equation}
 (cf.\ Sunyaev \& Titarchuk 1980). Thus, we use the geometry-independent 
parameters, $\alpha$ and $kT$, as the model parameters. 

For a specific geometry, we determine the actual optical depth of a source by 
a Monte Carlo method. The method is based on that of G\'orecki \& Wilczewski 
(1984). We have tested it in the range of $\tau$ from 0.5 to 5 against the 
corresponding results of Pozdnyakov, Sobol' \& Sunyaev (1983) and found 
excellent agreement. 

Fig.\ \ref{fig:tit} shows an example of the (excellent) agreement between the 
spectra from the Monte Carlo method (squares) and that of LZ87 (dashed curve). 
The low-energy spectral index is $\alpha=0.38$, which is imposed for both the 
solution of the Kompaneets equation and the Monte Carlo spectrum, and $kT=60$ 
keV. The optical depth used in the Kompaneets equation is 3.2, whereas the 
actual $\tau$ of a uniform sphere with a uniform distribution of seed photon 
sources is 3.0. This discrepancy in $\tau$ has a negligible practical 
importance as the derived optical depth is only representative for the 
astrophysical source of an unknown geometry. 

The method of LZ87 formally assumes that the Comptonizing plasma is optically 
thick. We find that $\tau\simgreat 2$ (for a sphere) is already sufficient for 
the validity of the method of LZ87. Fig.\ \ref{fig:par} in Section 
\ref{ss:par} shows a limiting case with $\tau=1.3$ ($\alpha=0.80$ and $kT=88$ 
keV for the LZ87 solution), when the assumption of the plasma being optically 
thick starts to become invalid. We see that then the method of LZ87 still 
gives an excellent description of the Monte Carlo results (at the same 
$\tau$), but it overestimates the value of $kT$, with the actual value from 
the simulations of $kT=61$ keV.

\begin{figure}
\begin{center}
\leavevmode
\epsfxsize=8.4cm \epsfbox{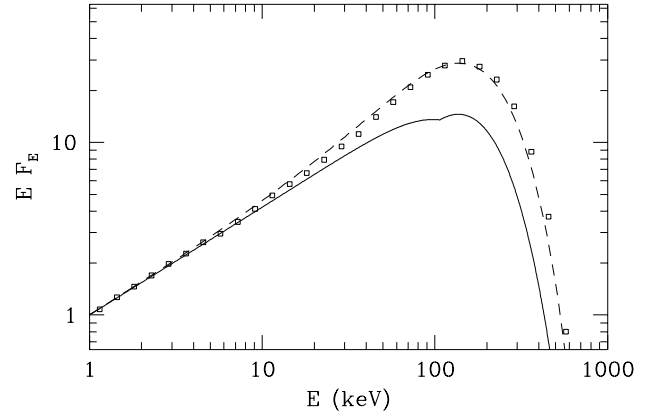}
\end{center}
\caption{Comparison of Comptonization spectra from a spherical plasma cloud 
with $kT=60$ keV, and $\tau$ such that the asymptotic low-energy index is 
$\alpha=0.38$. Squares represent Monte Carlo results, the solid curve 
corresponds to the model of Titarchuk (1994) and TM94, and the dashed curve 
corresponds to the solution of LZ87. 
 }
\label{fig:tit}
 \end{figure} 

Fig.\ \ref{fig:tit} also compares the Monte Carlo results (for $\tau\sim 3$, 
when the diffusion approximation applies) with those obtained using the 
solution of Titarchuk (1994; that solution is also published in TM94 and in a 
few other papers). Titarchuk (1994) also uses the Kompaneets equation, but 
with a set of relativistic corrections different than those in LZ87. We take 
$kT=60$ keV and $\alpha=0.38$ (corresponding to $\tau=3.1$) in the solution of 
Titarchuk (1994). We see that the spectrum of Titarchuk (1994) underestimates 
the actual spectrum in the range $E\sim kT$ by a factor of $\sim 2$. 
Furthermore, that solution gives an sharp kink around 100 keV, which is 
clearly unphysical. 

The kink appears due to an attempt [eq.\ (3) in TM94] to account for the 
relativistic high-energy cutoff steeper than that given by the Wien law 
[analogous to the diffusion-suppression factor of eq.\ (21) in LZ87]. For 
large values of $\alpha$, equation (3) in TM94 gives indeed a suppression with 
respect to the low-energy formula [eq.\ (2) in TM94]. However, the situation 
reverses at small values of $\alpha$, when the spectrum without the cutoff 
correction is already below the correct spectrum, and a hump appears in the 
spectrum above the kink, see Fig.\ \ref{fig:tit}. Concluding, the model of 
Titarchuk (1994) and TM94 may introduce large errors in fitting X\g\ data. 

\end{document}